# Incorporating patient-reported outcomes in dose-finding clinical trials with continuous patient enrollment

Anaïs Andrillon[a], Lucie Biard[a] and Shing M. Lee[b]*

[a]*INSERM U1153 Team ECSTRRA, Université Paris Cité, Paris, France;* [b]*Department of Biostatistics, Mailman School of Public Health, Columbia University, New York, NY, USA*

Dose-finding clinical trials in oncology aim to estimate the maximum tolerated dose (MTD), based on safety traditionally obtained from the clinician's perspective. While the collection of patient-reported outcomes (PROs) has been advocated to better inform treatment tolerability, there is a lack of guidance and methods on how to use PROs for dose assignments and recommendations. The PRO continual reassessment method (PRO-CRM) has been proposed to formally incorporate PROs to estimate the MTD, requiring complete follow-up of both clinician and patient toxicity information per dose cohort to assign the next cohort of patients. In this paper, we propose two extensions of the PRO-CRM, allowing continuous enrollment of patients and handling longer toxicity observation windows to capture late-onset or cumulative toxicities. The first method, the TITE-PRO-CRM, uses a weighted likelihood to include the partial follow-up information from PRO in estimating the MTD during and at the end of the trial. The second method, the TITE-CRM+PRO, uses clinician's information solely to inform dose assignments during the trial and incorporates PRO at the end of the trial for dose recommendation. Simulation studies show that the TITE-PRO-CRM performs similarly to the PRO-CRM in terms of dose recommendation and assignments during the trial while reducing trial duration. The TITE-CRM + PRO slightly underperforms compared to the TITE-PRO-CRM, but similar performance can be attained by requiring larger sample sizes. We also show that the proposed methods have similar performance under higher accrual rates, different toxicity hazards, and correlated time-to-clinician toxicity and time-to-patient toxicity data.

**Introduction**

The goal of dose-finding clinical trials is to identify the optimal dose or doses for further testing. In oncology, the optimal dose has conventionally been considered to be the maximum tolerated dose (MTD) which is selected based on safety to ensure that the drug or drug combination can be well tolerated. The information on tolerability has traditionally been obtained from the clinician perspective via the collection of adverse events using the National Cancer Institute Common Terminology Criteria of Adverse Events (NCI-CTCAE 2003). With recent studies suggesting under-reporting of symptoms by clinicians and lack of agreement between clinician and patient reported adverse events in both presence and severity (Basch et al. 2015), there has been increased interest in including patient-reported outcomes (PROs) in dose finding clinical trials to better inform drug tolerability (Basch 2016). A recently published survey of international stakeholders in cancer clinical trials showed that over 60% of responders agreed that PROs are helpful for identifying new toxicities and 100% agreed that PROs should be used to inform dose-escalation decisions (Lai-Kwon, Vanderbeek and Minchom 2022). Moreover, two recent studies evaluating the use of PRO in dose-finding trials suggested an increasing number of dose-finding clinical trials collecting PROs (Coleman 2021, Lai-Kwon et al. 2021).

The collection of PRO in early stage trials has been further advocated given the availability of validated PRO for the collection adverse events directly from patients, such as the patient-reported outcomes version of the CTCAE (PRO-CTCAE; (Dueck et al. 2015; Basch, Rogak and Dueck 2016)); more guidelines on the selection of PROs for dose finding trials (US FDA, 2021); and the impetus for PRO collection from project Optimus (US FDA Project Optimus). However, there is a lack of guidance on how to use PROs to inform dose assignments and dose recommendations for later phases of drug development. In a review of recently published dose-finding trials in Clinicaltrials.gov conducted by Lai-Kwon et al, the statistical analysis of PROs was mainly descriptive with 22% of manuscripts evaluating PRO by dose level, and only 16.7% using PROs to inform dose recommendation. Subjective use of PRO information based on descriptive summaries may lead to bias and lack of consistency. Thus, it is critical to propose approaches that can formally incorporate PRO in dose-finding clinical trials to inform drug tolerability. While there is extensive literature on methods for dose estimation, only one method has been proposed for formally incorporating PROs in the estimation of the tolerable dose, namely the patient-reported outcomes continual reassessment method (PRO-CRM; Lee, Lu and Cheng 2020). The method is an extension of the continual reassessment method (CRM) and defines the MTD as the dose that satisfies both a specified clinician toxicity threshold, as well as a specified patient toxicity threshold. It has been applied in practice for a trial of radiotherapy in patients with endometrial cancer (Wages et al. 2022). This method requires complete follow-up of both clinician and patient toxicity information per dose cohort to assign the next cohort of patients which is a limitation when continuous enrollment and longer toxicity observation windows to capture late-onset or cumulative toxicities are desired. To use the method enrollment needs to be suspended between cohorts and thus it is logistically challenging and trial duration may be extended. With the longer administration of novel cancer therapies, this is concerning. In that setting, it would be ideal to be able to include partial follow-up information to allow for continuous enrollment of patients. Methods for including partial follow-up have been proposed to estimate the dose associated with a given toxicity threshold. These include the time-to-event continual reassessment method (TITE-CRM; Cheung and Chappell (2000)) and the time-to-event Bayesian Optimal Interval Design (TITE-BOIN; Yuan et al. (2018)), among others.

In this paper, we propose methods for the use of PROs to inform drug tolerability in the setting of dose-finding clinical trials. First, we propose an extension of the PRO-CRM that uses a weighted likelihood to include the partial follow-up information from patients in the estimation of the model parameter during the trial like the TITE-CRM. The method named TITE-PRO-CRM is described in detail in Section 2. Given that this method would require the intense collection of PROs during the follow-up period, which is burdensome and may lead to missing data, we also propose methods using solely clinician information to inform dose assignment during the trial, and both patient and clinician at the end of the trial to inform dose recommendation. The results from simulation studies comparing the proposed methods to the PRO-CRM and TITE-CRM in the context of a motivating example are provided in Section 3. This is followed by a discussion in Section 4.

**Methods**

*Outcome Definition*

We consider a dose-finding clinical trial where patients are followed up for an observation window $[0, t*]$, defined for DLT observation and safety assessment. Using the same notation as Lee, Lu, and Cheng (2020), let $Y_c$ be the clinician-reported binary toxicity outcome, where $Y_c = 1$ indicates the occurrence of a DLT from the clinician perspective, and $Y_p$ be the patient-reported binary toxicity outcome, where $Y_p = 1$ indicates the occurrence of a DLT from the patient perspective. Let $D = \{d_1, \ldots, d_m\}$ be the set of dose levels of interest with $d_1 < \ldots < d_m$.

*Designs*

Based on separate DLT outcomes obtained from clinicians and patients, $Y_c$ and $Y_p$, we considered a marginal approach which allows for the specification of separate toxicity thresholds from the clinician ($\theta$) and the patient perspective ($\phi$). The target specification is in line with previous literature, and thus, more familiar for clinicians and easier to implement.

Moreover, simulation results show that the performance of the joint modeling and marginal modeling are similar except when the true MTD is the highest dose (Lee, Lu, and Cheng (2020)).

Based on the two separate DLT constraints from clinicians and patients, we define the optimal dose, $d^*$, as the minimum of the doses that satisfy the clinician and patient specified thresholds, that is

$$d^* = min\left\{argmax_{d_j \in D}\{P(Y_C = 1|d_j) \leq \theta\}, argmax_{d_j \in D}\{P(Y_P = 1|d_j) \leq \phi\}\right\}.$$

To estimate the probability of DLT from the clinician perspective, we consider an empiric working model

$$F_C(d_j) = u_j^\beta, \ d_j \in D, \ j \in \{1,\dots,m\},$$

where $\beta > 0$ is an unknown parameter to be estimated and $\{u_1,\dots,u_m\}$ are the scaled doses obtained by backward substitution from the dose skeleton as specified in the CRM. Similarly, we also consider an empiric working model to estimate the probability of DLT from the patient perspective

$$F_P(d_j) = v_j^\gamma, \ d_j \in D, \ j \in \{1,\dots,m\},$$

where $\gamma > 0$ is an unknown parameter to be estimated and $\{v_1,\dots,v_m\}$ are the scaled doses, which may differ from $\{u_1,\dots,u_m\}$.

To assign doses to newly enrolled patients based on partial DLT information from previously included patients and allow for continuous accrual of patients during the trial, we used an approach like the TITE-CRM (Cheung and Chappell 2000). Given the data accrued up to the first $i$ patients, we can estimate $\beta$ and $\gamma$ using weighted likelihoods based on clinician and patient outcomes separately. The weighted likelihood based on clinician-reported outcome is

defined as

$$L_i(\beta) = \prod_{k=1}^{i}\{w_C(k,i)\, u(k)^\beta\}^{y_C(k,i)} \{1 - w_C(k,i)\, u(k)^\beta\}^{1-y_C(k,i)},$$

where $y_C(k,i)$ is the indicator for clinician-reported toxicity for the $k^{th}$ patient, $w_C(k,i)$ the weight assigned to this observation just prior to the entry time of the $(i+1)^{th}$ patient, and $u(k)$ the scaled administered dose for the $k^{th}$ patient. Similarly, the weighted likelihood function based solely on patient-reported outcome is

$$L_i(\gamma) = \prod_{k=1}^{i}\{w_P(k,i)\, v(k)^\gamma\}^{y_P(k,i)} \{1 - w_P(k,i)\, v(k)^\gamma\}^{1-y_P(k,i)},$$

where $y_P(k,i)$ is the patient-reported toxicity indicator for patient $k^{th}$, $w_P(k,i)$ its weight, and $v(k)$ the scaled administered dose for the $k^{th}$ patient. In this paper, we use linear weight functions as this is suitable in most cases (Cheung and Chappell 2000). That is, suppose $[0, t^*]$ is the planned observation window for DLT and $l(k,i)$ is the available follow-up time for the $k^{th}$ patient prior to the entry of the $(i+1)^{th}$ patient: $w_C(k,i) = min\left(\frac{l(k,i)}{t^*}, 1\right)$, with $w_C(k,i) = 1$ if patient $k$ has experienced a clinician-reported DLT, and $w_P(k,i) = min\left(\frac{l(k,i)}{t^*}, 1\right)$, with $w_P(k,i) = 1$ if patient $k$ has experienced a patient-reported DLT. Parameters $\beta$ and $\gamma$ are estimated by Bayesian inference, and the posterior mean of each model parameter is computed and used as a plug-in estimate to obtain the dose with the estimated probability of clinician-reported DLT and the probability of patient-reported DLT closest to their respective specified targets. Then, the $(i+1)^{th}$ patient is assigned to,

$$\hat{d}_{i+1}^* = min\left\{argmin_{d_j \in D}\{|F_C(d_j)^{\hat{\beta}_i} - \theta|\}, argmin_{d_j \in D}\{|F_P(d_j)^{\hat{\gamma}_i} - \phi|\}\right\}.$$

*Dose-finding algorithms*

For the inclusion of PROs formally in dose assignments and recommendations, we proposed two approaches. In the first approach, named TITE-PRO-CRM, we use both the clinician-reported and patient-reported DLT information for dose assignments during the trial as well as the final dose recommendation. Thus, after each patient is enrolled, we apply the design specified above starting from the first patient. Using this approach, it is necessary to obtain updated toxicity information from both clinicians and patients during the conduct of the trial. In the second approach, namely TITE-CRM + PRO, we only use clinician DLT information to guide the dose assignment decisions during the trial and assign doses using TITE-CRM, and we use both the clinician and patient-reported DLT information for the estimation of the optimal dose, $d^*$, at the end of trial. That is, during the trial, the dose assigned to the $(i + 1)^{th}$ patient is,

$$\hat{d}^*_{i+1} = argmin_{d_j \in D} \left\{ |F_C(d_j)^{\widehat{\beta}_i} - \theta| \right\}.$$

When the total number of patients to be included, $N$, has been reached, we estimate the optimal dose, $d^*$, based on both clinician and patient outcomes, as

$$\widehat{d^*} = min \left\{ argmin_{d_j \in D} \left\{ |F_C(d_j)^{\widehat{\beta_N}} - \theta| \right\}, argmin_{d_j \in D} \left\{ |F_P(d_j)^{\widehat{\gamma_N}} - \phi| \right\} \right\}$$

where $\widehat{\beta_N}$ and $\widehat{\gamma_N}$ are the posterior means of $\beta$ and $\gamma$ after the outcomes for all $N$ patients have been observed.

**Simulation study**

*Motivating example*

We apply the proposed approaches to the same dose-finding trial settings that was used as a motivation example in Lee, Lu and Cheng (2020) for the PRO-CRM. Briefly, the dose-finding trial aimed to estimate the MTD of Bortezomib when administered in combination with CHOP + Rituximab (CHOP-R) (Leonard 2005). The trial evaluated 5 doses levels of

Bortezomib with CHOP + Rituximab (CHOP-R) and it was designed using CRM with a target probability of DLT from the clinician perspective of 0.25 with a sample size of 18 patients.

*Simulation setting*

We conducted a simulation study to assess the performance of the proposed TITE-PRO-CRM and TITE-CRM + PRO designs, in comparison to the TITE-CRM and the PRO-CRM. The TITE-CRM was selected as a comparator to evaluate the effect of not including PROs in the dose-finding trial, although the objectives differ. The PRO-CRM was selected to investigate the effect of using partial information during the conduct of the trial. According to the setting of the motivating example, we considered five dose levels $d_1 < \ldots < d_5$, a target probability of clinician defined DLT by t* of 0.25 ($\theta = 0.25$) and a target probability of DLT from the patient perspective by t* of 0.35 ($\phi = 0.35$) (Lee, Lu, and Cheng 2020). For our simulations, we assumed an observation window of t* = 6 weeks, and a starting dose at dose level 1. The cohort size was 1 patient, and no dose skipping was allowed during dose-escalation. We evaluated sample sizes of 18, 30, and 40 patients to assess the effect of sample size. The dose skeletons and the prior variance of the parameters $\beta$ and $\gamma$ were jointly calibrated using the concepts of the least informative prior variance and indifference intervals developed by Lee and Cheung, which provides a systematic approach for selecting a skeleton with optimized operating characteristics (Lee and Cheung 2011). The skeletons were calibrated separately for the clinician and patient working models. For the clinician constraint alone, given five dose levels, a target DLT rate of 0.25, and assuming the MTD was dose level 3, the dose skeleton was $u = (0.08, 0.16, 0.25, 0.35, 0.46)$, for N=18 and 40 and (0.06, 0.14, 0.25, 0.38, 0.50) for N=30. For the patient constraint alone, given five dose levels, a target DLT rate of 0.35, and assuming the MTD is dose level 3, the skeleton was $v =$

(0.13, 0.23, 0.35, 0.47, 0.58) for N=18 and 30, and (0.10, 0.21, 0.35, 0.49, 0.61), for N=40. A normal prior was specified for $\beta$ and $\gamma$, with mean 0 and the least informative prior standard deviation which was 0.522 for N=18, and 40 and 0.627 for N=30 for the clinician constraint and 0.59 for N=18 and 30, and 0.69 for the patient constraint.

To evaluate the performance of the proposed method, we examined the seven clinician-patient toxicity probability scenarios used previously in the paper by Lee, Lu, and Cheng (2020). In the first four scenarios, the clinician MTD coincides with $d^*$, the MTD which incorporates both clinician and patient-reported outcomes. We intentionally selected scenarios in which the MTD is dose level 3 or higher when the clinician MTD and patient MTD coincided because we expected the method to be conservative and do well when the MTD is one of the lower dose levels. In scenarios 5 to 7, the MTD based solely on clinician DLT differs from $d^*$, and thus it is expected that the recommended dose for the TITE-CRM would differ from the methods incorporating PROs.

For data generation, patients' failure times were obtained under a Weibull model considering time-constant and time-varying toxicity hazards. The shape parameter was set to 1, 0.3, and 3, to simulate time-constant, decreasing or increasing hazard respectively. In each scenario, the scale parameter was then obtained based on the shape parameter and the toxicity probability at time $t^*$. Furthermore, to evaluate the performance of the method in the case of correlated time-to-clinician toxicity and time-to-patient toxicity data, we used Clayton's model for generating time-to-event following Yuan and Yin's approach (Yuan and Yin 2009). We set the correlation term to 0.1 and 0.9 to induce a low and high positive dependence between the clinician and patient time-to-events. As a sensitivity analysis, we also evaluated the method under different accrual rates by considering an expected number of 2 and 4 patient arrivals per observation window.

For each scenario, we performed 10,000 simulations and we estimated: the probability of selecting each dose level as final MTD, the probability of correct selection (PCS), the average number of patients treated at a dose above the true MTD (No. OD) and treated at the true MTD (No. MTD) during the trial, the average number of patients that experienced a clinician DLT (No. Clin DLT) and a patient-reported DLT (No. Pat DLT) during the trial, and the total trial duration time in weeks.

**Results**

Tables 1 and 2 summarize the performance of the proposed designs for the seven scenarios, when the total sample size is 18, the patient accrual is set at two patients per time window, with constant hazard for failure times and under a low level of correlation between time-to-clinician toxicity and time-to-patient toxicity. Tables 3 and 4 are for N=40 patients with the same specifications. First, the TITE-PRO-CRM maintained the operating characteristics of the PRO-CRM for all scenarios. In Scenarios 1 to 4, when the MTD based solely on clinician outcome coincided with the MTD incorporating both clinician and patient-reported outcomes, the TITE-PRO-CRM like the PRO-CRM tended to be more conservative and reduced both the allocation and the recommendation of over-toxic doses compared to the TITE-CRM (Tables 1 and 3). Thus, for Scenario 3 where the MTD is located at the highest dose level the TITE-PRO-CRM under performs compared to the TITE-CRM (PCS of 41% vs 67% for N=18). This improved with an increased sample size, and the PCS for the TITE-PRO-CRM reached 54% with N=30 patients and 60% with N=40.

When the MTD based on both clinician and patient-reported outcomes is lower than the MTD based solely on clinician outcome (Scenarios 5 to 7; Tables 2 and 4), the TITE-CRM as expected recommended higher dose levels with considerably higher probabilities, given that it does not take into account the patient constraint. The TITE-PRO-CRM was able to account for both patient and clinician constraints and recommended the correct dose in more than 56% of cases with a sample size of 18 regardless of the distance between the clinician and patient MTD. In terms of safety, as expected, the number of overdose and the number of observed patient and clinician DLT were much lower with the TITE-PRO-CRM compared to the TITE-CRM for Scenarios 5 to 7. PCS improved to 60%-80% across all scenarios with a sample size of 40 (Table 3 and 4).

The performance of the TITE-PRO-CRM was similar in terms of percent of correct selection even under a faster patient accrual and different types of toxicity hazard (Figure 1). Furthermore, the average number of patients treated at over-toxic doses, the average number of patients that experienced clinician and patient reported DLT during the trial were slightly lower in case of decreasing hazard (i.e., early onset toxicities) for all scenarios (data not shown). Under correlated time-to-clinician toxicity and time-to-patient toxicity data, the TITE-PRO-CRM tended to be more aggressive and recommended higher doses like the TITE-CRM, improving the performance under Scenario 3 by increasing the PCS from 41% to 50% when the MTD is located at the highest dose level. However, the performance worsened (63% versus 58%) when the MTD was dose level 3 in Scenario 1.

The second proposed approach, the TITE-PRO+CRM, using only clinician DLT information to guide the dose assignment decision during the trial and using both the clinician and patient-reported DLT information for the final estimation of the optimal dose, tended to underperform compared to the TITE-PRO-CRM in terms of the probability of correct selection, while being more conservative in recommending over-toxic doses (Table 1 and 2). However, during the trial the TITE-PRO + CRM was more aggressive in terms of dose assignments, allocating more patients to toxic doses on average and exposing more patients to patient and clinician DLT compared to the TITE-PRO-CRM given that it did not consider the PRO information during the trial. The performance of the method improved with larger sample sizes indicating that larger sample sizes could be used to achieve similar performance to the TITE-PRO-CRM. As expected, both methods reduced trial duration: with an accrual rate of 2 patients per DLT window, the trial duration was almost halved by using the TITE-PRO-CRM and TITE-CRM + PRO (average duration of a trial of 57 weeks versus 108 weeks with the PRO-CRM).

**Discussion**

In this paper, we propose methods that incorporate patient-reported outcomes in dose-finding clinical trials with the aim of identifying the optimal dose of a drug which satisfies constraints both from the clinician and the patient perspective. Unlike previous proposed methods, these methods do not require complete observations, thus allowing for continuous enrollment, shorter trial duration, and formal inclusion of the patient perspective in the final dose recommendation. The first method, the TITE-PRO-CRM, is an extension of the PRO-CRM (Lee, Lu and Cheng 2020) whereby the model likelihood is sequentially weighted during the trial, to account for partial follow-ups at the time of dose assignments, like the TITE-CRM (Cheung and Chappell 2000). The method uses both clinician-reported and patient-reported outcomes for each dose assignment and the final dose estimation, and performs similarly to the PRO-CRM while reducing trial duration and similarly to the TITE-CRM in scenarios where the MTD for the clinician and patient constraints are the same except when the MTD is the highest dose. In such cases a higher sample size may be desired to compensate for the conservativeness of the method. Decrease in trial duration was generally consistent with the accrual rate with close to 50%-decrease on average in case of an accrual of 2 patients per observation window.

Additionally, we propose an alternative approach, the TITE-CRM + PRO, incorporating PRO data only at the end of the trial after full accrual, for MTD estimation, while sequential dose assignments during the trial rely on the standard TITE-CRM using clinician-reported outcomes. While formally incorporating PROs for the estimation of the optimal dose, this method does not require real-time monitoring of PRO data during the trial. Thus, it may be more practical for operational reasons, given the burden that repeated collection of PROs can be to patients and sites, and the likelihood of missing data. For a given sample size, the TITE-CRM + PRO slightly underperforms compared to the TITE-PRO-CRM which uses PRO throughout the trial. Relying solely on clinician-reported toxicity during the trial results on average in slightly more aggressive dose assignments, over-estimated patient-reported toxicity rates particularly at lower dose levels only, and eventually more conservative final dose recommendation. Nevertheless, TITE-CRM + PRO performance improves with larger sample size. Thus, when using this method, one may opt for a larger sample size to compensate for the loss in the probability of correct selection. In addition, we also evaluate using isotonic regression at the end of the trial, as proposed in the BOIN design (Liu and Yuan 2015), to estimate the dose based on the patient constraint, and it performs similarly (data not shown).

Patient-reported outcomes can provide a unique perspective in terms of treatment tolerability and have become crucial, notably in oncology. The recent Friends of Cancer white paper Optimizing Dosing in Oncology Drug Development has advocated for the collection of PROs at the earliest clinical stages of drug development to inform dose optimization and tolerability of treatment. There is also been a growing interest in incorporating patients' perspectives, including from regulatory purposes (Basch and Yap 2021; Lai-Kwon, Vanderbeek and Minchom 2022; Wages et al. 2022; Retzer et al. 2022). However, guidelines for how to use the PRO data collected for the recommendation of the optimal dose are currently lacking and much needed. In this paper, two approaches to incorporate PRO in dose-finding clinical trials early in drug development both during the trial for dose assignment or only after full accrual for dose recommendation. These methods also allow for continuous accrual enhancing the applicability of the design in practice, avoiding the need to wait for complete observations before dose assignments.

**Declaration of interest statement**

The authors declare no potential conflict of interests.

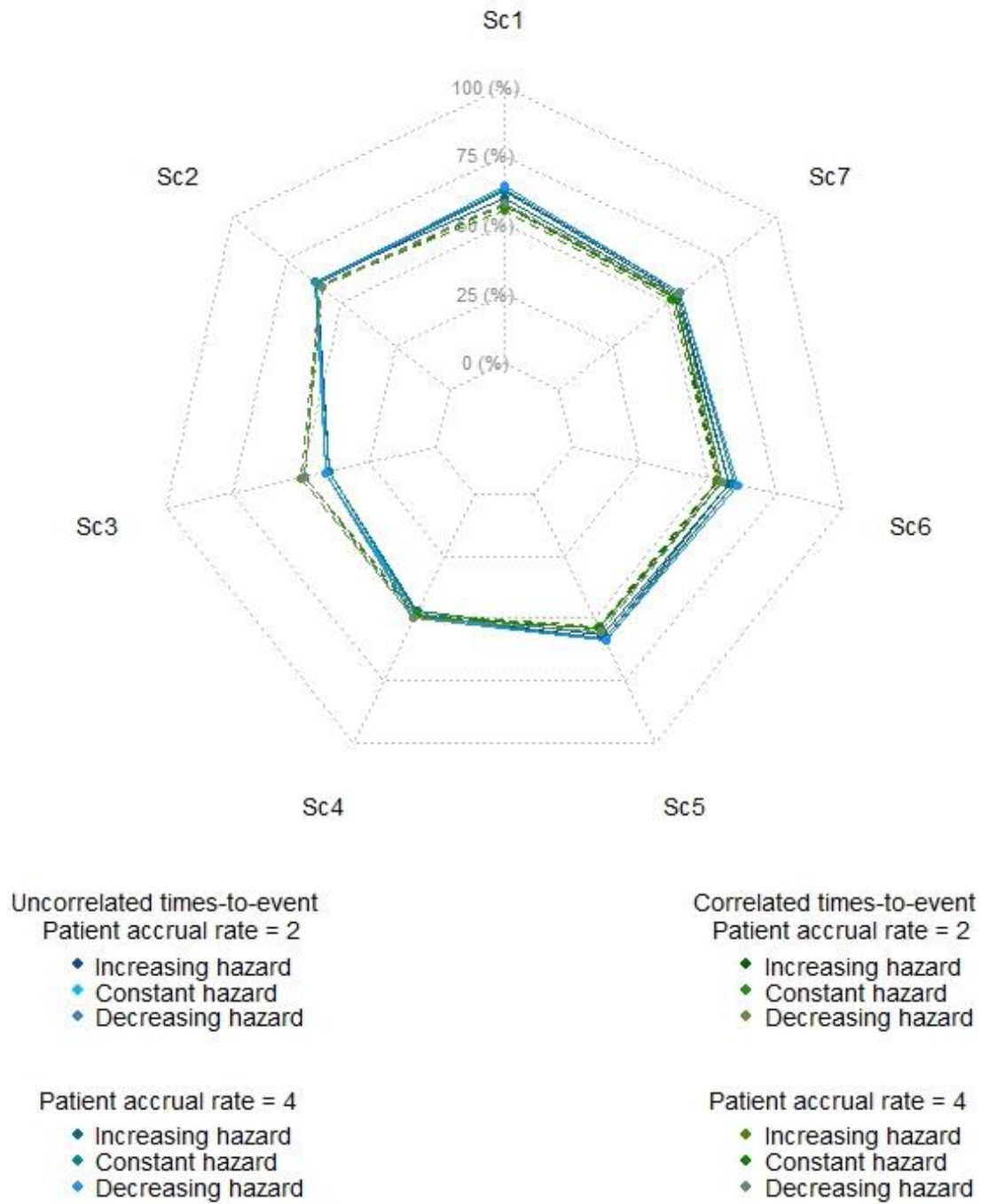

Figure 1. Percent of correct selection for Scenarios 1 to 7 with the TITE-PRO-CRM according to the level of correlation between time-to-events, patient arrival rates by observation window and the shape of hazard with N=18 patients per trial.

# Tables

Table 1. Simulation results for Scenarios 1 to 4 under low correlation between time-to-events with N=18, a constant hazard of toxicity and an accrual rate of 2 patients expected per observation window.

|  | % selection by dose level | | | | | No. OD | No. Pat DLT | No. Clin DLT | No. MTD |
|---|---|---|---|---|---|---|---|---|---|
|  | **1** | **2** | **3** | **4** | **5** | | | | |
| *Scenario 1* | | | | | | | | | |
| Probability of DLT by Clinician | 0.05 | 0.05 | **0.25** | 0.4 | 0.55 | | | | |
| Probability of DLT by Patient | 0.17 | 0.18 | **0.35** | 0.5 | 0.65 | | | | |
| TITE-CRM | 0 | 15 | **61** | 23 | 2 | 5.10 | 6.31 | 4.34 | 8.02 |
| PRO-CRM | 2 | 28 | **63** | 7 | 0 | 1.57 | 5.18 | 3.10 | 8.32 |
| TITE-PRO-CRM | 2 | 28 | **63** | 7 | 0 | 1.77 | 5.20 | 3.13 | 8.06 |
| TITE-CRM + PRO | 4 | 36 | **55** | 5 | 0 | 5.10 | 6.31 | 4.34 | 8.02 |
| *Scenario 2* | | | | | | | | | |
| Probability of DLT by Clinician | 0.05 | **0.25** | 0.4 | 0.55 | 0.7 | | | | |
| Probability of DLT by Patient | 0.1 | 0.15 | **0.35** | 0.5 | 0.65 | | | | |
| TITE-CRM | 11 | **60** | 27 | 2 | 0 | 7.19 | 4.32 | 5.33 | 8.19 |
| PRO-CRM | 11 | **62** | 27 | 1 | 0 | 5.70 | 3.82 | 4.86 | 9.42 |
| TITE-PRO-CRM | 10 | **62** | 27 | 1 | 0 | 5.97 | 3.91 | 4.92 | 9.09 |
| TITE-CRM + PRO | 12 | **65** | 23 | 1 | 0 | 7.19 | 4.32 | 5.33 | 8.19 |
| *Scenario 3* | | | | | | | | | |
| Probability of DLT by Clinician | 0.01 | 0.02 | 0.05 | 0.1 | **0.25** | | | | |
| Probability of DLT by Patient | 0.04 | 0.09 | 0.17 | 0.2 | **0.35** | | | | |
| TITE-CRM | 0 | 0 | 2 | 30 | **67** | 0 | 4.17 | 2.39 | 5.99 |
| PRO-CRM | 0 | 1 | 11 | 48 | **40** | 0 | 3.63 | 1.84 | 3.38 |
| TITE-PRO-CRM | 0 | 1 | 11 | 48 | **41** | 0 | 3.60 | 1.80 | 3.10 |
| TITE-CRM + PRO | 0 | 0 | 10 | 50 | **40** | 0 | 4.17 | 2.39 | 5.99 |
| *Scenario 4* | | | | | | | | | |
| Probability of DLT by Clinician | 0.02 | 0.05 | 0.1 | **0.25** | 0.4 | | | | |
| Probability of DLT by Patient | 0.09 | 0.17 | 0.2 | **0.35** | 0.5 | | | | |
| TITE-CRM | 0 | 1 | 24 | **57** | 18 | 1.94 | 5.28 | 3.43 | 8.16 |
| PRO-CRM | 0 | 6 | 42 | **48** | 4 | 0.48 | 4.36 | 2.50 | 5.54 |
| TITE-PRO-CRM | 0 | 6 | 42 | **48** | 4 | 0.42 | 4.37 | 2.51 | 5.7 |
| TITE-CRM + PRO | 0 | 6 | 47 | **44** | 3 | 1.94 | 5.28 | 3.43 | 8.16 |

Table 2. Simulation results for Scenarios 5 to 7 under low correlation between time-to-events with N=18, a constant hazard of toxicity and an accrual rate of two patients expected per observation window.

|  | % selection by dose level | | | | | No. OD | No. Pat DLT | No. Clin DLT | No. MTD |
|---|---|---|---|---|---|---|---|---|---|
|  | 1 | 2 | 3 | 4 | 5 | | | | |
| *Scenario 5* | | | | | | | | | |
| Probability of DLT by Clinician | 0.05 | 0.1 | 0.16 | **0.25** | 0.4 | | | | |
| Probability of DLT by Patient | 0.05 | 0.2 | **0.35** | 0.5 | 0.65 | | | | |
| TITE-CRM | 0 | 7 | **33** | 46 | 15 | 8.47 | 7.08 | 3.58 | 5.62 |
| PRO-CRM | 1 | 25 | **59** | 15 | 0 | 2.65 | 5.42 | 2.60 | 8.04 |
| TITE-PRO-CRM | 1 | 25 | **59** | 15 | 1 | 2.84 | 5.46 | 2.62 | 7.95 |
| TITE-CRM + PRO | 4 | 32 | **51** | 12 | 0 | 8.47 | 7.08 | 3.58 | 5.62 |
| *Scenario 6* | | | | | | | | | |
| Probability of DLT by Clinician | 0.05 | 0.18 | 0.2 | **0.25** | 0.4 | | | | |
| Probability of DLT by Patient | 0.17 | **0.35** | 0.5 | 0.65 | 0.8 | | | | |
| TITE-CRM | 2 | **17** | 32 | 36 | 12 | 12.56 | 9.22 | 3.85 | 3.94 |
| PRO-CRM | 17 | **60** | 22 | 1 | 0 | 4.86 | 6.42 | 2.81 | 8.99 |
| TITE-PRO-CRM | 16 | **60** | 22 | 1 | 0 | 5.17 | 6.48 | 2.83 | 8.64 |
| TITE-CRM + PRO | 33 | **51** | 15 | 1 | 0 | 12.56 | 9.22 | 3.85 | 3.94 |
| *Scenario 7* | | | | | | | | | |
| Probability of DLT by Clinician | 0.01 | 0.05 | 0.1 | 0.16 | **0.25** | | | | |
| Probability of DLT by Patient | 0.04 | 0.05 | 0.2 | **0.35** | 0.5 | | | | |
| TITE-CRM | 0 | 1 | 11 | **40** | 48 | 4.21 | 5.60 | 2.70 | 7.59 |
| PRO-CRM | 0 | 2 | 32 | **56** | 10 | 1.11 | 4.48 | 2.16 | 6.86 |
| TITE-PRO-CRM | 0 | 2 | 32 | **56** | 10 | 0.98 | 4.49 | 2.15 | 7.07 |
| TITE-CRM + PRO | 0 | 5 | 37 | **49** | 9 | 4.21 | 5.62 | 2.70 | 7.59 |

Table 3. Simulation results for Scenarios 1 to 4 under low correlation between time-to-events with N=40, a constant hazard of toxicity and an accrual rate of two patients expected per observation window.

|  | % selection by dose level |  |  |  |  | No. OD | No. Pat DLT | No. Clin DLT | No. MTD |
|---|---|---|---|---|---|---|---|---|---|
|  | 1 | 2 | 3 | 4 | 5 |  |  |  |  |
| *Scenario 1* |  |  |  |  |  |  |  |  |  |
| Probability of DLT by Clinician | 0.05 | 0.05 | **0.25** | 0.4 | 0.55 |  |  |  |  |
| Probability of DLT by Patient | 0.17 | 0.18 | **0.35** | 0.5 | 0.65 |  |  |  |  |
| TITE-CRM | 0 | 8 | **76** | 16 | 0 | 9.47 | 14.30 | 10.04 | 23.11 |
| PRO-CRM | 0 | 18 | **78** | 3 | 0 | 2.51 | 12.11 | 7.68 | 23.78 |
| TITE-PRO-CRM | 0 | 18 | **78** | 3 | 0 | 2.69 | 12.10 | 7.66 | 23.36 |
| TITE-CRM + PRO | 0 | 25 | **72** | 2 | 0 | 9.47 | 14.30 | 10.04 | 23.11 |
| *Scenario 2* |  |  |  |  |  |  |  |  |  |
| Probability of DLT by Clinician | 0.05 | **0.25** | 0.4 | 0.55 | 0.7 |  |  |  |  |
| Probability of DLT by Patient | 0.1 | 0.15 | **0.35** | 0.5 | 0.65 |  |  |  |  |
| TITE-CRM | 6 | **76** | 18 | 0 | 0 | 12.21 | 8.57 | 11.26 | 23.30 |
| PRO-CRM | 6 | **75** | 19 | 0 | 0 | 10.54 | 8.03 | 10.77 | 24.79 |
| TITE-PRO-CRM | 6 | **75** | 19 | 0 | 0 | 10.91 | 8.11 | 10.82 | 24.31 |
| TITE-CRM + PRO | 6 | **78** | 16 | 0 | 0 | 12.21 | 8.57 | 11.26 | 23.30 |
| *Scenario 3* |  |  |  |  |  |  |  |  |  |
| Probability of DLT by Clinician | 0.01 | 0.02 | 0.05 | 0.1 | **0.25** |  |  |  |  |
| Probability of DLT by Patient | 0.04 | 0.09 | 0.17 | 0.2 | **0.35** |  |  |  |  |
| TITE-CRM | 0 | 0 | 0 | 18 | **82** | 0 | 11.06 | 7.06 | 22.31 |
| PRO-CRM | 0 | 0 | 3 | 37 | **60** | 0 | 9.61 | 5.57 | 14.08 |
| TITE-PRO-CRM | 0 | 0 | 3 | 37 | **60** | 0 | 9.55 | 5.52 | 13.74 |
| TITE-CRM + PRO | 0 | 0 | 1 | 41 | **58** | 0 | 11.06 | 7.06 | 22.31 |
| *Scenario 4* |  |  |  |  |  |  |  |  |  |
| Probability of DLT by Clinician | 0.02 | 0.05 | 0.1 | **0.25** | 0.4 |  |  |  |  |
| Probability of DLT by Patient | 0.09 | 0.17 | 0.2 | **0.35** | 0.5 |  |  |  |  |
| TITE-CRM | 0 | 0 | 15 | **73** | 13 | 5.19 | 12.82 | 8.75 | 22.38 |
| PRO-CRM | 0 | 2 | 30 | **66** | 2 | 1.03 | 10.77 | 6.71 | 18.02 |
| TITE-PRO-CRM | 0 | 2 | 30 | **66** | 2 | 0.98 | 10.76 | 6.68 | 17.97 |
| TITE-CRM + PRO | 0 | 0 | 36 | **62** | 2 | 5.19 | 12.82 | 8.75 | 22.38 |

Table 4. Simulation results for Scenarios 5 to 7 under low correlation between time-to-events with N=40, a constant hazard of toxicity and an accrual rate of two patients expected per observation window.

|  | % selection by dose level | | | | | No. OD | No. Pat DLT | No. Clin DLT | No. MTD |
|---|---|---|---|---|---|---|---|---|---|
|  | 1 | 2 | 3 | 4 | 5 | | | | |
| *Scenario 5* | | | | | | | | | |
| Probability of DLT by Clinician | 0.05 | 0.1 | 0.16 | **0.25** | 0.4 | | | | |
| Probability of DLT by Patient | 0.05 | 0.2 | **0.35** | 0.5 | 0.65 | | | | |
| TITE-CRM | 0 | 2 | **29** | 58 | 11 | 22.61 | 17.19 | 8.77 | 12.60 |
| PRO-CRM | 0 | 15 | **74** | 11 | 0 | 5.31 | 12.89 | 6.13 | 22.88 |
| TITE-PRO-CRM | 0 | 15 | **74** | 11 | 0 | 5.51 | 12.92 | 6.15 | 22.75 |
| TITE-CRM + PRO | 1 | 22 | **68** | 10 | 0 | 22.61 | 17.19 | 8.77 | 12.60 |
| *Scenario 6* | | | | | | | | | |
| Probability of DLT by Clinician | 0.05 | 0.18 | 0.2 | **0.25** | 0.4 | | | | |
| Probability of DLT by Patient | 0.17 | **0.35** | 0.5 | 0.65 | 0.8 | | | | |
| TITE-CRM | 0 | **13** | 34 | 44 | 9 | 30.97 | 21.69 | 9.09 | 7.32 |
| PRO-CRM | 9 | **76** | 15 | 0 | 0 | 8.87 | 14.29 | 6.56 | 24.35 |
| TITE-PRO-CRM | 9 | **76** | 15 | 0 | 0 | 9.18 | 14.34 | 6.56 | 23.94 |
| TITE-CRM + PRO | 25 | **64** | 11 | 0 | 0 | 30.97 | 21.69 | 9.09 | 7.32 |
| *Scenario 7* | | | | | | | | | |
| Probability of DLT by Clinician | 0.01 | 0.05 | 0.1 | 0.16 | **0.25** | | | | |
| Probability of DLT by Patient | 0.04 | 0.05 | 0.2 | **0.35** | 0.5 | | | | |
| TITE-CRM | 0 | 0 | 3 | **36** | 61 | 16.08 | 14.91 | 7.23 | 16.14 |
| PRO-CRM | 0 | 0 | 19 | **72** | 9 | 3.02 | 11.6 | 5.51 | 20.98 |
| TITE-PRO-CRM | 0 | 0 | 18 | **73** | 9 | 2.90 | 11.58 | 5.50 | 21.08 |
| TITE-CRM + PRO | 0 | 1 | 26 | **66** | 7 | 16.08 | 14.91 | 7.23 | 16.14 |